\documentclass[useAMS,usenatbib]{mnras}
\usepackage{amsmath,amstext,amsgen,amsbsy,amsopn,amsfonts,theorem}

\usepackage[pdftex]{graphicx}
\usepackage{xspace}
\usepackage{amsmath}
\usepackage{framed}
\usepackage{txfonts}
\usepackage{epstopdf}
\usepackage{gensymb}
\usepackage{color}

\usepackage[normalem]{ulem}

\makeatletter
\def\mr@ignsp#1 {\ifx\:#1\@empty\else #1\expandafter\mr@ignsp\fi}%
\newcommand{\multiref}[1]{\begingroup
\xdef\mr@no@sparg{\expandafter\mr@ignsp#1 \: }%
\def\mr@comma{}%
\@for\mr@refs:=\mr@no@sparg\do{\mr@comma\def\mr@comma{,}\ref{\mr@refs}}%
\endgroup}
\makeatother

\newcommand{\hypref}[2]{\ifx\href\asklfhas #2\else\href{#1}{#2}\fi}
\newcommand{\Secref}[1]{Section~\multiref{#1}}
\newcommand{\secref}[1]{Sec.~\multiref{#1}}

\newcommand{\Tabref}[1]{Table~\multiref{#1}}

\newcommand{\Figref}[1]{Figure~\multiref{#1}}
\newcommand{\figref}[1]{Fig.~\multiref{#1}}
\renewcommand{\eqref}[1]{(\multiref{#1})}

\title[Sloshing of the Galactic halo]{Detection of the LMC-induced sloshing of the Galactic halo}

\author[D. Erkal et al.]{
\parbox{\textwidth}{
  \Large
  Denis Erkal$^{1}$\thanks{d.erkal@surrey.ac.uk},
  Alis J. Deason$^{2,3}$,
  Vasily Belokurov$^4$,
  Xiang-Xiang Xue$^5$,
  Sergey E. Koposov$^{6,4,7}$,
  Sarah A. Bird$^{8,9}$,
  Chao Liu$^8$,
  Iulia T. Simion$^9$,
  Chengqun Yang$^9$,
  Lan Zhang$^5$,
  Gang Zhao$^5$
}
\vspace{0.4cm}
\\
\parbox{\textwidth}{
 $^1$Department of Physics, University of Surrey, Guildford GU2 7XH, UK \\
  $^{2}$Institute for Computational Cosmology, Department of Physics, University of Durham, South Road, Durham DH1 3LE, UK\\
$^{3}$Centre for Extragalactic Astronomy, Department of Physics, University of Durham, South Road, Durham DH1 3LE, UK\\
$^{4}$Institute of Astronomy, Madingley Rd, Cambridge, CB3 0HA \\
$^{5}$CAS Key Laboratory of Optical Astronomy, National Astronomical Observatories, Beijing 100101, China \\
$^6$Institute for Astronomy, University of Edinburgh, Royal Observatory, Blackford Hill, Edinburgh EH9 3HJ, UK \\
$^7$McWilliams Center for Cosmology, Carnegie Mellon University, 5000 Forbes Ave, 15213 \\
$^8$Key lab of Space Astronomy and Technology, National Astronomical Observatories, Beijing 100101, China \\
$^9$Shanghai Astronomical Observatory, 80 Nandan Road, Shanghai 200030, China \\
}
}

\begin{document}

\label{firstpage}

\maketitle

\begin{abstract}
A wealth of recent studies have shown that the LMC is likely massive, with a halo mass $>10^{11} M_\odot$. One consequence of having such a nearby and massive neighbour is that the inner Milky Way is expected to be accelerated with respect to our Galaxy's outskirts (beyond $\sim 30$ kpc). In this work we compile a sample of $\sim 500$ stars with radial velocities in the distant stellar halo, $r_{\rm GC}> 50$ kpc, to test this hypothesis. These stars span a large fraction of the sky and thus give a global view of the stellar halo. We find that stars in the Southern hemisphere are on average blueshifted, while stars in the North are redshifted, consistent with the expected, mostly downwards acceleration of the inner halo due to the LMC. We compare these results with simulations and find the signal is consistent with the infall of a $1.5\times10^{11} M_\odot$ LMC. We cross-match our stellar sample with \textit{Gaia} DR2 and find that the mean proper motions are not yet precise enough to discern the LMC's effect. Our results show that the outer Milky Way is significantly out of equilibrium and that the LMC has a substantial effect on our Galaxy.
\end{abstract}


\section{Introduction}

The Large Magellanic Cloud (LMC) is the most luminous satellite of the Milky Way (MW) and has been known about since antiquity \cite[e.g.][]{alsufi}. Given its stellar mass, abundance matching predicts that the LMC occupies a halo which had a peak mass of $\sim2\times10^{11} M_\odot$ \citep[e.g.][]{moster_etal_2013,behroozi_etal_2013}. Since the LMC is likely on its first approach to the Milky Way \citep[e.g.][]{kallivayalil_2006_LMCSMC_orbit,besla_etal_2007,kallivayalil_etal_2013}, its dark matter halo should only recently have been tidally deformed by our Galaxy and should still be relatively close to the LMC. This LMC mass is comparable to the enclosed mass of the Milky Way within $50$ kpc, $\sim4\times10^{11} M_\odot$ \citep[e.g.][]{Wang_etal_2020}. 

Several studies have now confirmed that the dynamical influence of the LMC is consistent with its expected dark matter halo. \cite{kallivayalil_etal_2013} showed that an LMC mass $>10^{11} M_\odot$ is needed in order for the Small Magellanic Cloud (SMC) to have been bound to the LMC. \cite{lmc_sats} extended this argument for all of the (then-known) Magellanic satellites and found an LMC mass $>1.24\times10^{11} M_\odot$ is needed if they were originally bound to the LMC. \cite{penarrubia_lmc} studied the timing argument of the MW with M31, as well as the local Hubble flow and found an LMC mass of $\sim(2.5\pm0.9)\times10^{11} M_\odot$ was needed. Finally, \cite{orphan_modelling,sgr_tango} showed that the Orphan and Sagittarius streams can only be explained by including the presence of a $\sim(1.4\pm0.3)\times10^{11} M_\odot$ LMC. 

\cite{gomez_et_al_2015} showed that another exciting consequence of having such a massive LMC is that our Galaxy will move in response to the LMC's approach. Indeed, the stream fits of \cite{orphan_modelling,sgr_tango} both showed that allowing the Milky Way to respond to the LMC gave the best fits to the observational data. Based on their predicted response of the Milky Way to the LMC, \cite{orphan_modelling} argued that as the LMC tugs on the Milky Way during its approach, the inner part of our Galaxy will respond coherently due to the short orbital timescales in this region. As a result, they predicted that the inner $\sim30$ kpc of our Galaxy should have a bulk velocity relative to the outskirts of the Milky Way. Due to the recent orbit of the LMC, this bulk velocity should primarily be in the downwards (i.e. $-z$) direction. 

Subsequent simulations \citep[e.g.][]{lmc_wake,petersen_penarrubia,Erkal2020,cunningham_2020,Garavito-Camargo_etal_2020} showed that this picture was roughly correct although the response of the Milky Way's outer halo is more subtle. These works made predictions for the kinematic signatures in the stellar halo of the Milky Way and showed that the dominant signal will be in the radial velocity and in the proper motion in Galactic latitude, $\mu_b$. The prediction for the radial velocity is a negative radial velocity in the South and a positive radial velocity in the North due to the downwards motion of the inner Milky Way with respect to its outskirts. Similarly, the proper motion is predicted to be upwards, i.e. $\mu_b > 0$, throughout the distant stellar halo. \cite{Erkal2020} attempted to detect this effect by looking at the mean 3D velocity of 33 globular clusters and dwarf galaxies with Galactocentric radii larger than 30 kpc. They found a significant upwards mean velocity, consistent with the picture that we are roughly moving downwards compared to the outer halo. 

This ``sloshing" in the outskirts of our Galaxy has a number of important implications. First, this means that the outskirts of our Galaxy are not in equilibrium which must be taken into account when measuring the Milky Way mass beyond $r>30$ kpc to avoid biases \citep[][]{Erkal2020}. Second, this also implies that the dark matter halo of the Milky Way has also been dramatically deformed \citep[e.g.][]{lmc_wake,petersen_penarrubia,Garavito-Camargo_etal_2020}. Detecting these changes would be a stunning confirmation of the dark matter paradigm. 

In this work, we present the first detection of the bulk motion of the Milky Way's stellar halo and show it is consistent with the expected effect of the LMC. In order to do this, we construct a large sample of stars ($\sim500$) with measured radial velocities in the distant, $r_{\rm GC}>50$ kpc, stellar halo. In \Secref{sec:data} we will describe this data set. In \Secref{sec:sims} we compare the properties of our stellar sample with simulations of the Milky Way stellar halo which include the effect of the LMC, and find a good agreement. We discuss the implications and limitations of this result in \Secref{sec:discussion} and conclude in \Secref{sec:conclusions}.

\section{Data} \label{sec:data}

Since the close approach of the LMC effectively decouples the inner $\sim30$ kpc of the Milky Way from its outskirts \citep[e.g.][]{orphan_modelling,lmc_wake,petersen_penarrubia,Erkal2020}, a rich set of kinematic features are predicted in the outer stellar halo. In this work we will mainly focus on the radial velocity signature but we will also explore the proper motions in \Secref{sec:pms}. In this section, we will describe the sample of stars that we have selected from the literature.

Given that the LMC does not produce any strong effect on the inner Milky Way halo, we focus on stars with Galactocentric radii beyond $50$ kpc where the impact of the LMC's in-fall should be prominent. Our sample is made up of a variety of different stellar types. First, we have blue horizontal branch (BHB) stars and blue stragglers (BS) from \cite{Deason2012_BHBs,Xue2008,pisces_LMC_wake}. Next we have a sample of K-giants from \cite{Xue2014,yang_etal_2019}. Finally, we have a sample of RR Lyrae from \cite{Cohen+2017}. Due to the rapid dropoff in the number of observed stars at large radii, we truncate the sample at $105$ kpc so that we can make a meaningful comparison with our simulated stellar halo. See \Tabref{tab:properties} for a summary of our stellar sample. 

In order to get rid of contaminants, we cross-match our sample to sources in \textit{Gaia} DR2 within 1 arcsecond \citep{GaiaDR2}. While all of the K-giant stars are in the \textit{Gaia} DR2 catalogue, a number of BHBs and RR Lyrae do not appear. For the stars with astrometry (i.e. proper motions and parallaxes), we compute the uncertainties on the total speeds by Monte-Carlo sampling the errors on their observables 10,000 times, including the covariance in proper motion. The Sun is placed at a distance of 8.122 kpc from the Galactic center \citep{2018A&A...615L..15G}, moving with a velocity of (11.1,\,245,\,7.3) km/s motivated by \cite{schoenrich_etal_2010,bovy_etal_vcirc}. We remove stars with $\varpi/\sigma_\varpi > 3$ and those with $v_{\rm tot}-\sigma_{v_{\rm tot}} > 500$ km/s. Note that we do not remove any of the stars which do not have astrometry in \textit{Gaia} DR2.

\begin{figure}
\centering
\includegraphics[width=0.45\textwidth]{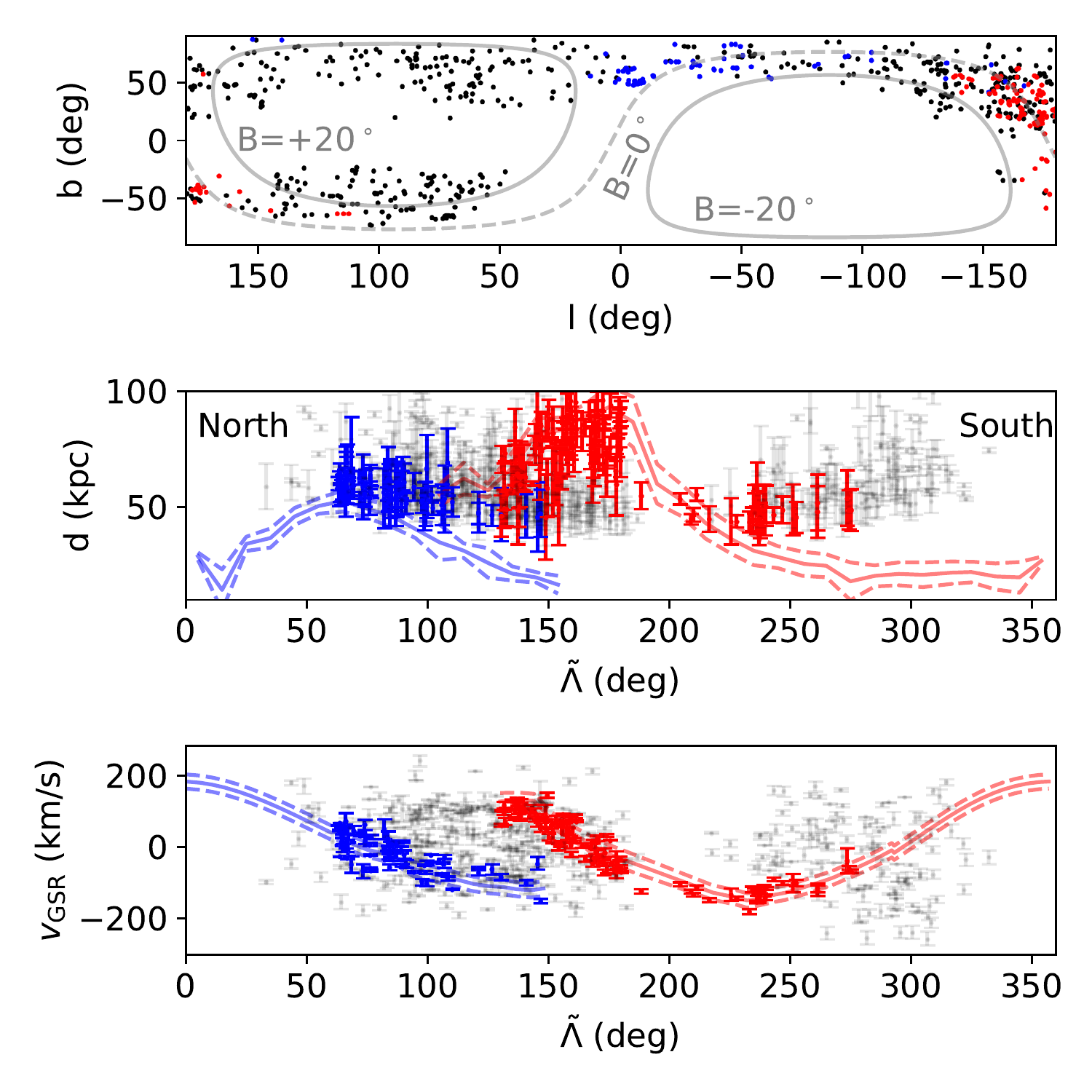}
\caption{Selection of Sagittarius stars. \textit{Top panel} shows the on-sky location of stars in Galactic coordinates. We also show lines of constant latitude ($B=-20^\circ,0^\circ,20^\circ$) in the Sgr stream coordinates of \protect\cite{belokurov_sgr_precess}. \textit{Middle} and \textit{bottom} panels show the heliocentric distance and radial velocity relative to the Galactic standard of rest versus the longitude in Sgr stream coordinates, $\tilde{\Lambda}$. In each panel the black points show stars that we associate with the stellar halo, while the blue (red) points show stars in the leading (trailing) arms of the Sagittarius stream. In the middle and bottom panel, the red (blue) lines show the observed trend and uncertainty for the leading (trailing) arms of Sagittarius \protect\citep[][]{belokurov_sgr_precess,sgr_tango}. Note that all of the stars shown in this figure have passed our astrometric cuts described in \protect\secref{sec:data}}   \label{fig:sgr_selection}
\end{figure}

Finally, we remove stars associated with the Sagittarius (Sgr) stream. We select these stars based on their position on the sky, their distance, and their radial velocity. We use the Sgr coordinates from \cite{belokurov_sgr_precess} and make a cut in latitude, $|B| < 20^\circ$. For the distance, we use the results of \cite{Hernitschek2017} who provide the mean and spread (i.e. standard deviation) of the distances along the Sgr stream. We require Sgr stars to be within $3\sigma$ of the mean distance track. For the radial velocity, we combine the radial velocity compilation of \cite{belokurov_sgr_precess} with the radial velocity spline in \cite{sgr_tango}. For the uncertainty, we convolve the uncertainties in \cite{belokurov_sgr_precess} with $20$ km/s based on the Sgr velocity dispersion reported in \cite{Gibbons2017}. As with the distance, we require the Sgr stars to be within $3\sigma$ of the observed radial velocity track. The stars which pass all three of these cuts are assigned to the leading or trailing arm of Sgr while those that fail any of the cuts are assigned to the stellar halo. \Figref{fig:sgr_selection} shows the sample of stars which pass our astrometric cuts, classified by whether they are assigned to the stellar halo (black), leading arm of Sgr (blue), or trailing arm of Sgr (red).

With the final, clean sample, we show the radial velocity versus Galactic latitude in \Figref{fig:b_vs_vgsr}. This figure shows a clear difference in the mean radial velocity between the Northern and Southern hemispheres. In the South, the mean radial velocity is significantly negative while in the North it is slightly positive. The red line with error bars shows the mean radial velocity computed in $10^\circ$ bins. The light blue points and dashed-blue line show stellar halo particles from our fiducial simulation in which the Milky Way stellar halo is evolved in the presence of a $1.5\times10^{11} M_\odot$ LMC. We describe this simulation in \Secref{sec:sims}. To test the robustness of this result, in \Tabref{tab:properties} we show that this negative mean in the South and positive mean in the North is present in each of the individual data sets used in this work. We note that the radial velocity depends on the location on the sky \citep[see e.g. Fig. 1 of][]{Erkal2020} and thus it is not surprising that each sample has a different mean. 

\begin{table*}
\begin{centering}
\begin{tabular}{|c|c|c|c|c|c|c|}
\hline
Sample & Star type & $50 < r_{\rm GC}/({\rm kpc}) < 105$ & Astrometric cuts & Sgr cuts & $\langle v_{\rm GSR} \rangle_{b<0^\circ}$ (km/s) & $\langle v_{\rm GSR} \rangle_{b>0^\circ}$ (km/s) \\ \hline
\cite{Xue2014} & K-giants & 280 & 275 & 189 & $-9.4\pm15.1$ & $16.1\pm7.1$\\
\cite{yang_etal_2019} & K-giants & 301 & 171 & 101 & $-6.3\pm21.8$ & $4.1\pm9.9$\\
\cite{Xue2008} & BHB/BS & 123 & 113 & 99 & $-30.8\pm21.3$ & $11.3\pm9.8$ \\
\cite{Cohen+2017} & RR Lyrae & 111 & 88 & 86 & $-66.7\pm16.8$ & $2.9\pm10.5$ \\
\cite{Deason2012_BHBs} & BHB/BS & 23 & 22 & 9 & $-$ & $-$ \\
\cite{pisces_LMC_wake} & BHB & 8 & 8 & 8 & $-$ & $-$ \\ \\
\textbf{Total}   &   & 846  & 677  & 492  & $-27.2\pm9.1$ & $10.7\pm4.4$ \\
\end{tabular}
\caption{Summary of stellar sample used in this work. Note that the rows are organized according to the number of stars which pass the final cuts. Column 1 gives the reference for each sample and Column 2 gives the star type. Columns 3,4,5 show the number of stars which pass each of the cuts in this work. Note that the cuts are cumulative, i.e. Column 5 shows the number of stars which pass all of the cuts. Column 6,7 show the mean radial velocity in the Southern and Northern hemispheres. We have omitted this for the samples of \protect\cite{Deason2012_BHBs} and \protect\cite{pisces_LMC_wake} due to their small size.}
\label{tab:properties}
\end{centering}
\end{table*}

\section{Comparison with simulations} \label{sec:sims}

We now compare the observed radial velocities with the results of a suite of simulations of the Milky Way stellar halo in the presence of the LMC. Several of these simulations have already been presented in \cite{pisces_LMC_wake,Erkal2020} and for completeness we will briefly describe them again here. 

In order to account for response of the Milky Way due to the LMC, we model both systems as individual particles sourcing their host potentials. The Milky Way is modelled with a potential similar to \texttt{MWPotential2014} from \cite{bovy_galpy}: an NFW halo \citep{nfw_1997} with a mass of $8\times10^{11} M_\odot$, a scale radius of $16$ kpc, and a concentration of 15.3, a Miyamoto-Nagai disk \citep{mn_disk} with a mass of $6.8\times10^{10} M_\odot$, a scale height of $0.28$ kpc, and a scale length of $3$ kpc, and a Hernquist bulge \citep{hernquist_1990} with a mass of $5\times10^9 M_\odot$ and a scale radius of 0.5 kpc. We account for the dynamical friction of the Milky Way on the LMC using the results of \cite{jethwa_lmc_sats}. As described in \cite{pisces_LMC_wake}, the Milky Way stellar halo is initialized with $10^7$ tracer particles with an anisotropy of $\beta \sim0.5$ and a density profile of $\rho \propto r^{-3.5}$ at large radii using \textsc{agama} \citep{agama}. The LMC is modelled as a Hernquist profile with masses of $[0.2,0.5,1,1.5,2,2.5,3]\times10^{11} M_\odot$. For each LMC mass, we fix the scale radius so that the circular velocity at 8.7 kpc matches the observed value of 91.7 km/s \citep{vandermarel_lmc}. The LMC is initialized based on its present day distance, proper motion, and line-of-sight radial velocity \citep{pietrzynski_lmc_dist,kallivayalil_etal_2013,vandermarel_lmc_rv} and rewound for 5 Gyr. At this time, the Milky Way stellar halo is initialized and evolved to the present. 

We note that while these simulations capture many important aspects of the Milky Way-LMC interaction, we model the potentials of both galaxies as rigid and thus we are neglecting their tidal deformations \citep[e.g.][]{lmc_wake,petersen_penarrubia,Garavito-Camargo_etal_2020}. Based on the similarity between the predictions of \cite{Erkal2020}, which use identical simulations to those in this work, and the predictions of \cite{lmc_wake} we believe that our models are capturing the salient parts of how the Milky Way stellar halo is affected by the LMC.

\begin{figure}
\centering
\includegraphics[width=0.45\textwidth]{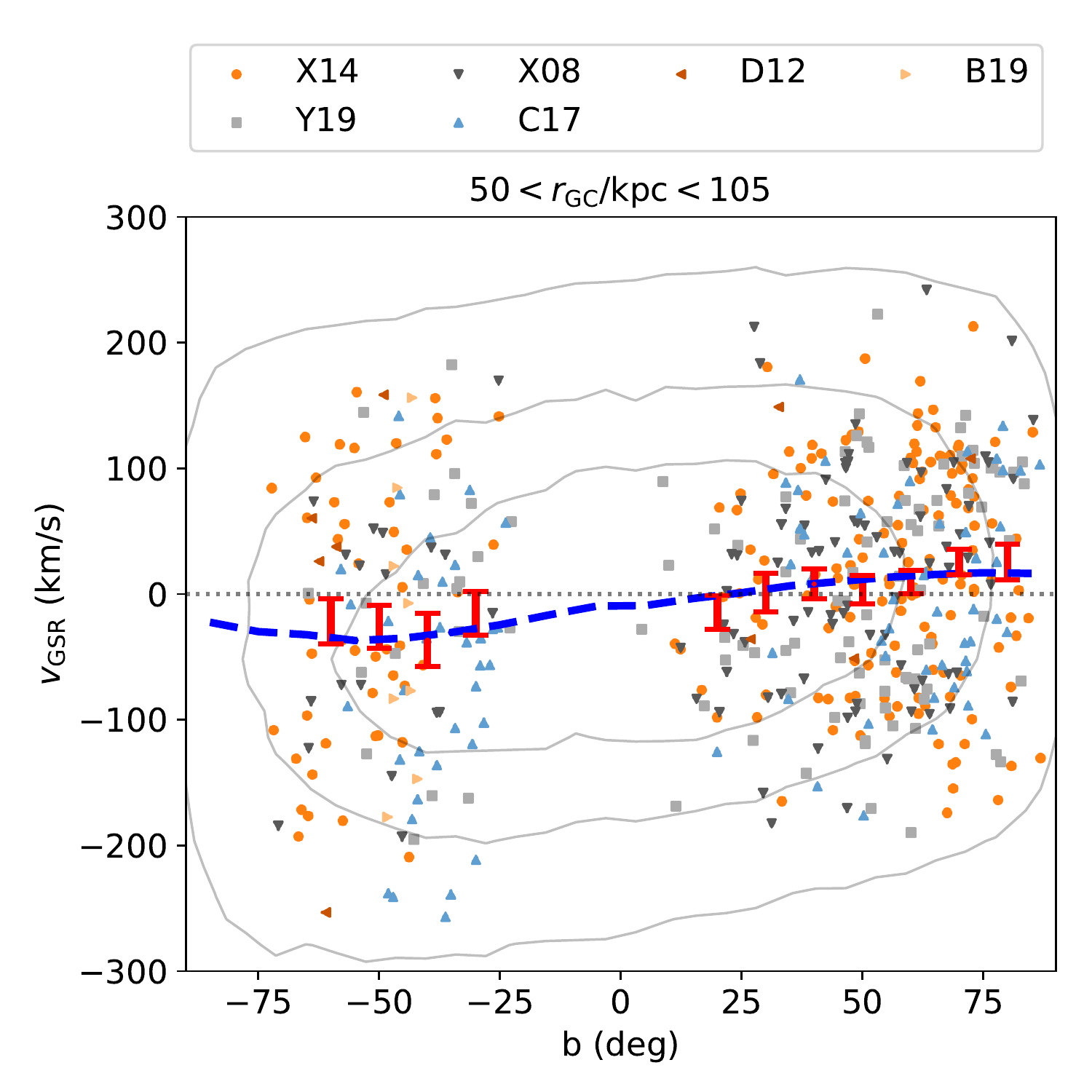}
\caption{Galacocentric radial velocity versus Galactic latitude for halo stars with $50 < r_{\rm GC} < 105$ kpc. The grey contours show the distribution in a simulated Milky Way stellar evolved in the presence of a $1.5\times10^{11} M_\odot$ LMC. The larger colored symbols show the observations from a variety of sources: X14 \protect\citep{Xue2014}, Y19 \protect\citep{yang_etal_2019}, X08 \protect\citep{Xue2008}, C17 \protect\cite{Cohen+2017}, D12 \protect\citep{deason_etal_2012}, B19 \protect\citep{pisces_LMC_wake}. The dashed blue line shows the mean radial velocity in the simulated stellar halo. The red points and error bars show the mean and error on the mean for the data. The dotted black line shows the $v_{\rm GSR}=0$ km/s for reference. The data follows the same overall trend as the simulation with a negative mean in the Southern hemisphere and a slightly positive mean in the Northern hemisphere. Note that we do not include the stars classified as Sgr members in this figure.}   \label{fig:b_vs_vgsr}
\end{figure}

In \Figref{fig:b_vs_vgsr} we show a contour plot of the radial velocity versus Galactic latitude for our fiducial simulation with an LMC mass of $1.5\times10^{11} M_\odot$. This fiducial LMC mass is selected since it produces a signal consistent with the observed radial velocities and matches previous measurements of the LMC mass \citep[e.g.][]{orphan_modelling,sgr_tango}. This figure shows contours for all simulated particles with Galactocentric radii between $50$ and $105$ kpc to match our sample. For this comparison, we have removed the particles with $l<0^\circ, b<0^\circ$ since we have very few stars in this quadrant on the sky (see top panel of \figref{fig:sgr_selection}). Note that as a result we have undersampled the simulated particles in the North by a factor of 2 so that the North and South are evenly sampled in the figure. The dashed-blue line shows the mean radial velocity computed in $10^\circ$ bins. Similar to the data, this shows a negative radial velocity in the Southern hemisphere and a positive radial velocity in the Northern hemisphere. 

\begin{figure}
\centering
\includegraphics[width=0.45\textwidth]{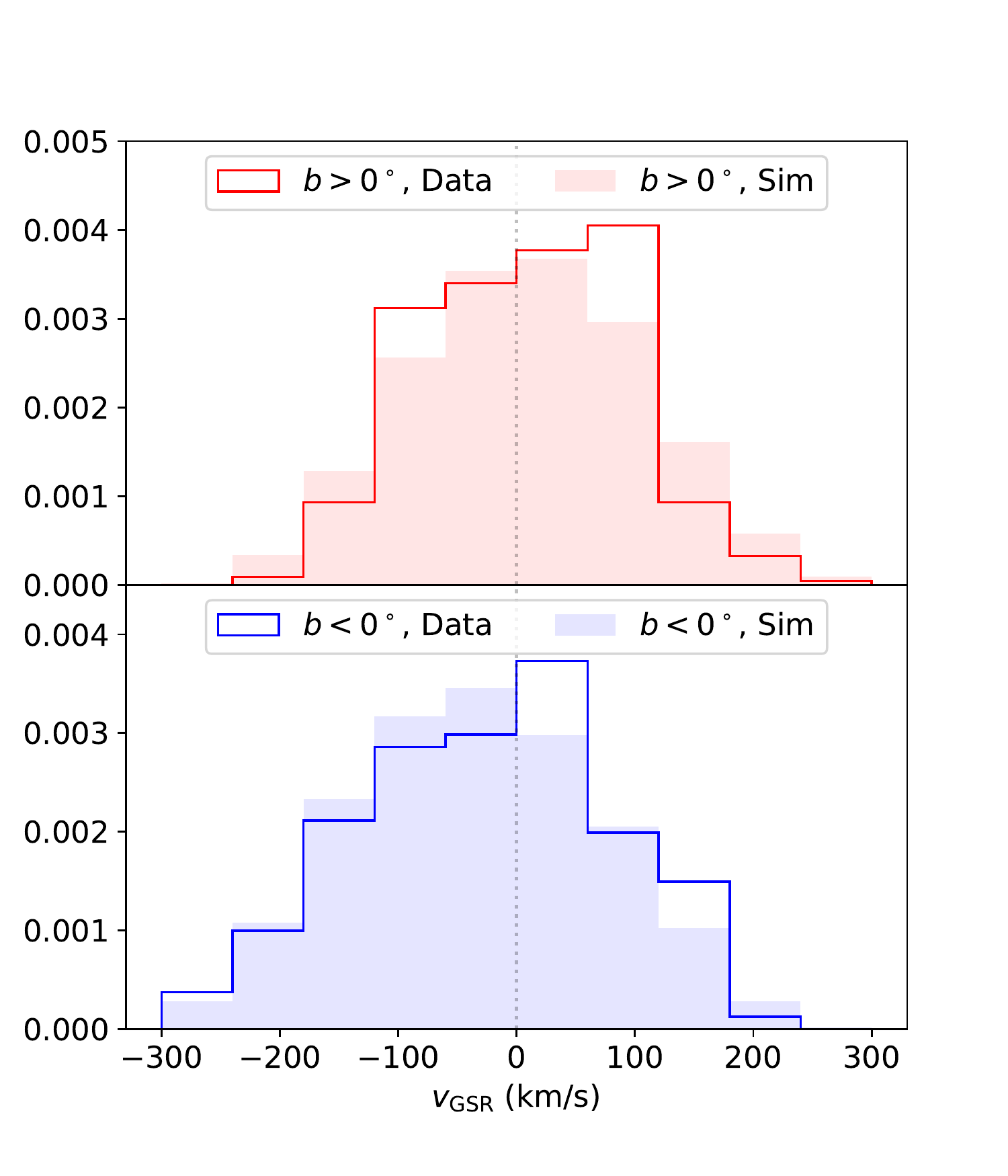}
\caption{Distribution of radial velocities for halo stars, corrected for the Sun's motion, in the Northern and Southern hemisphere. \textit{Top panel} shows the distribution of radial velocities in the Northern hemisphere. The unfilled histogram shows the distribution of observed radial velocities while the shaded histogram shows the radial velocity distribution in our fiducial simulation with an LMC mass of $1.5\times10^{11} M_\odot$. The dotted vertical line shows 0 km/s for reference. Both distributions are shifted to positive radial velocities. \textit{Bottom panel} shows the distribution of radial velocities in the Southern hemisphere. These are shifted slightly towards negative radial velocities. Note that the color scheme is chosen to emphasize the redshift in the North and blueshift in the South.}  \label{fig:vgsr_dist}
\end{figure}

Next, we compute the distribution of radial velocities in the Northern and Southern hemisphere in the data and compare this with our fiducial model. This is shown in \Figref{fig:vgsr_dist}. In order to make a fair comparison, for each star in our sample, we select the 100 closest particles (in 3D distance) in the fiducial simulation. As expected from \Figref{fig:b_vs_vgsr}, this shows that the radial velocity in the Southern hemisphere has been shifted to negative velocities while the radial velocities in the North have been shifted to slightly positive velocities. Overall, the simulations show a similar distribution of velocities to the data.  Note that there appears to be some substructure in the Southern hemisphere with an excess of stars with radial velocity $0<v_{\rm GSR} < 50$ km/s. We will discuss this further in \Secref{sec:substructure}.

\begin{figure}
\centering
\includegraphics[width=0.45\textwidth]{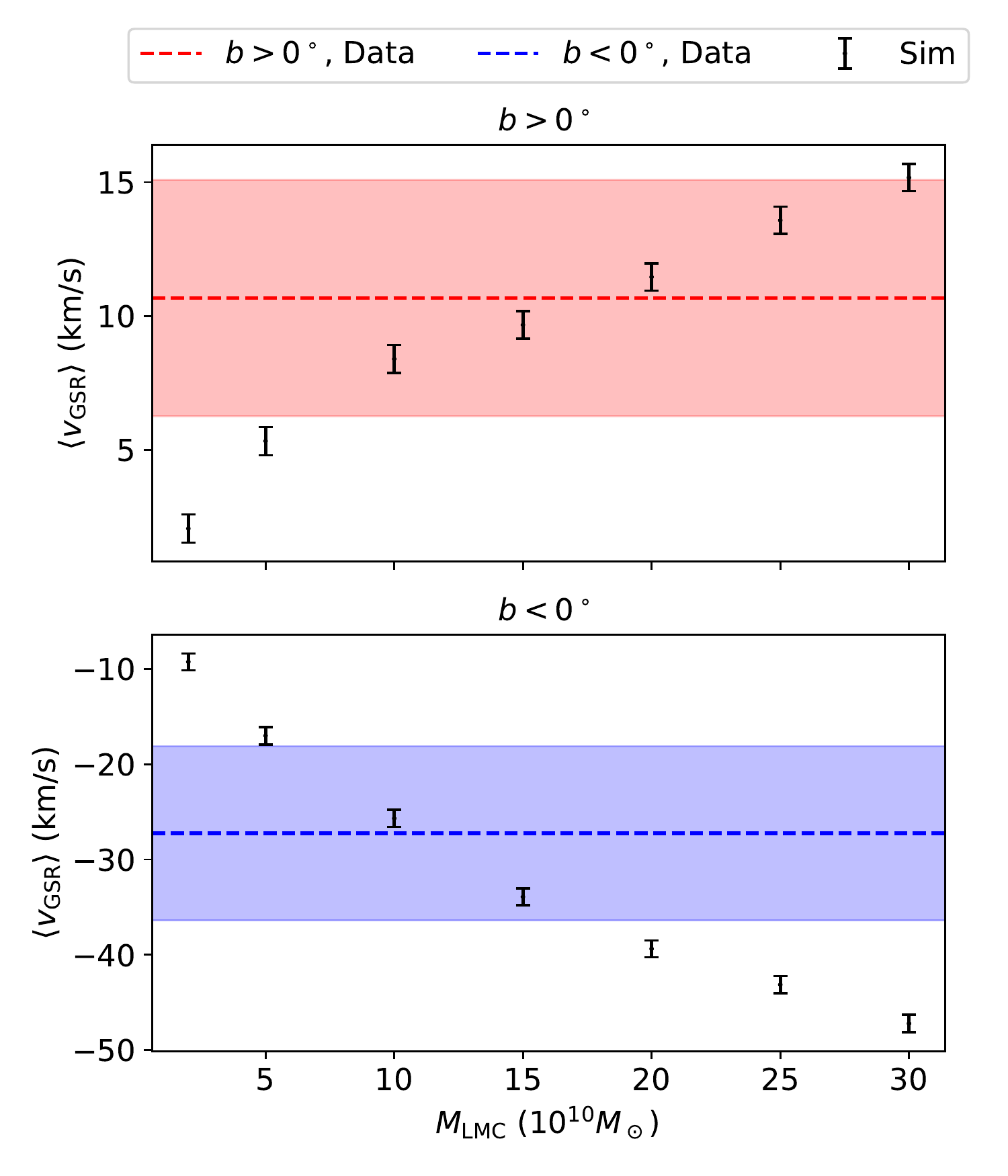}
\caption{Comparison of the mean radial velocity in observations and simulations. The \textit{top panel} and \textit{bottom panel} show the mean radial velocity in the Northern and Southern hemisphere respectively. In each panel, the mean and uncertainty of the observations are shown with the dashed-line and shaded region. The black error bars show the mean radial velocity in the simulated stellar haloes in the presence of difference LMC masses. As in the models, the data shows a positive/negative radial velocity in the North and South, consistent with the expected reflex motion of the inner Milky Way. }  \label{fig:vgsr_vs_mlmc}
\end{figure}

In \Figref{fig:vgsr_vs_mlmc} we fit a Gaussian to the radial velocities in the Northern and Southern hemispheres separately. We also perform the same fit on the particles in the simulation matched to the observed sample for the 7 different LMC masses under consideration. Note that the sample in the simulation is $100$ times larger which explains the substantially smaller error bars. The top panel and bottom panels show the mean radial velocity in the North and South respectively. Overall, the mean radial velocity exhibits a dipole where the mean in the North is redshifted, i.e. moving away from us, while the mean radial velocity in the South is blueshifted, i.e. moving towards us. This is consistent with the picture in which the LMC accelerates the inner part of the Milky Way with respect to its outskirts \citep[e.g.][]{orphan_modelling,lmc_wake,petersen_penarrubia,Erkal2020}. A closer look shows that bulk motion in the Southern hemisphere has a larger velocity shift than in the Northern hemisphere. This velocity signature is in good agreement with the simulations and previous predictions \citep[e.g.][]{lmc_wake,petersen_penarrubia,cunningham_2020}. 

\Figref{fig:vgsr_vs_mlmc} also shows that the observed shifts in radial velocity require a fairly substantial LMC. In the North, the signal is consistent with an LMC mass between $\sim0.5-3\times10^{11} M_\odot$ while in the South it is consistent with a $\sim 0.5-2\times10^{11} M_\odot$ LMC. With larger data sets from upcoming surveys such as the WHT Enhanced Area Velocity Explorer \citep[WEAVE,][]{weave}, the Dark Energy Spectroscopic Instrument \citep[DESI,][]{desi} and the 4-metre Multi-Object Spectroscopic Telescope \citep[4MOST,][]{4most}, it will be possible to measure this shift much more precisely and thus also better constrain the mass of the LMC and its effect on the Milky Way. 

\begin{figure*}
\centering
\includegraphics[width=0.99\textwidth]{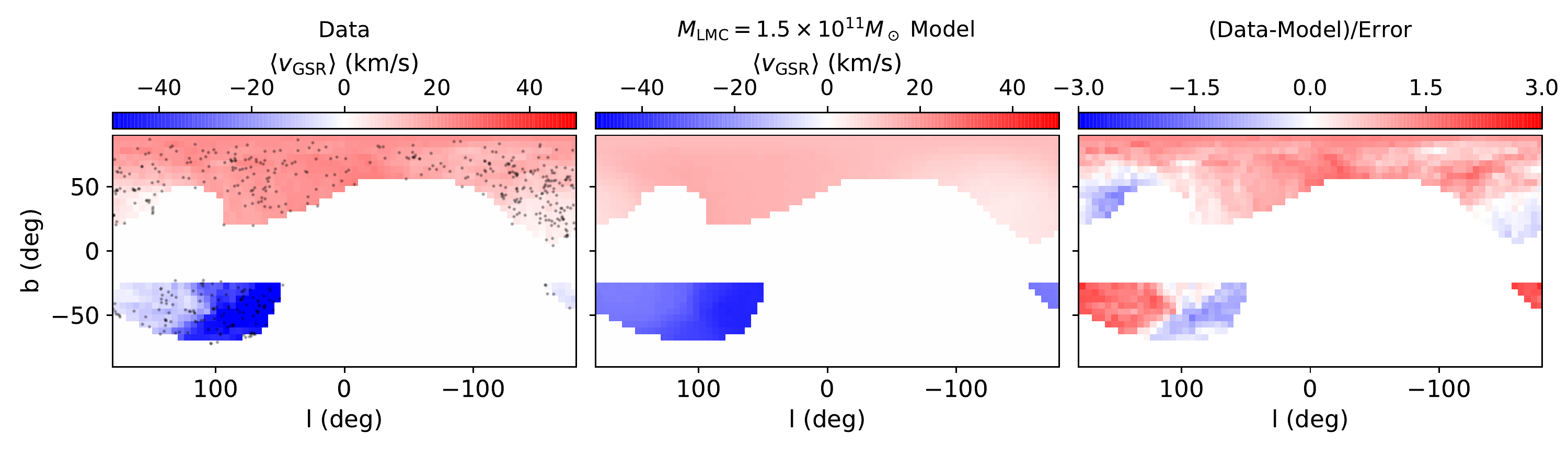}
\caption{Mean radial velocity over the footprint of the stellar sample considered in this work. For each pixel on the sky, we compute the mean velocity of the nearest half of the stellar sample in the same hemisphere. Thus this figure is effectively averaged on scales corresponding to roughly half the sample's extent in the North and South. \textit{Left panel} shows the mean radial velocity for our stellar sample. The black points show the location of the stellar sample used in this work. \textit{Middle panel} shows the mean radial velocity for the simulated stellar halo. Note that the sample of stars in the simulated stellar halo is 100 times larger which makes the map smoother. \textit{Right panel} shows the difference between the data and the model normalized by the uncertainty in the data. This shows that the mean radial velocity over sky broadly matches the expected signature of a $1.5\times10^{11} M_\odot$ LMC.}  \label{fig:vgsr_on_sky}
\end{figure*}

Finally, we explore how the radial velocity shift varies with position on the sky in \Figref{fig:vgsr_on_sky}. Since the stellar halo velocity dispersion is substantial, $\sigma \sim 100$ km/s, we need to average over many stars in order to measure the shift. In this figure, for each pixel on the sky, we fit a Gaussian to the radial velocities of the nearest half of the sample (as measured by angular distance on the sky). This choice is made so that the uncertainty on the mean for each pixel is roughly constant. Thus, this figure is effectively smoothed on very large scales corresponding to half of the sample's extent in the Northern and Southern hemispheres and we stress that pixels are strongly correlated on scales smaller than this. 

The left panel of \Figref{fig:vgsr_on_sky} shows the average radial velocity of the data. Note that the footprint comes from a convex hull placed around the sample. In the South, we defined this hull using Galactic coordinates while in the North we defined the hull using equatorial coordinates (RA/Dec). These choices were made to avoid empty regions with no nearby stars. The middle panel shows the average radial velocity in the fiducial simulation with an LMC mass of $1.5\times10^{11} M_\odot$. Note that the sample in the simulation is 100 times larger than the observed sample and hence the averages are much smoother. The right panel shows the difference between the data and the model, normalized by the uncertainty on the mean from the data. Overall, the model is broadly similar to the data although there are regions in the South $(l>100^\circ)$ where the model is a poor match. 

\section{Discussion} \label{sec:discussion}

\subsection{Proper motion signal} \label{sec:pms}

\begin{figure}
\centering
\includegraphics[width=0.45\textwidth]{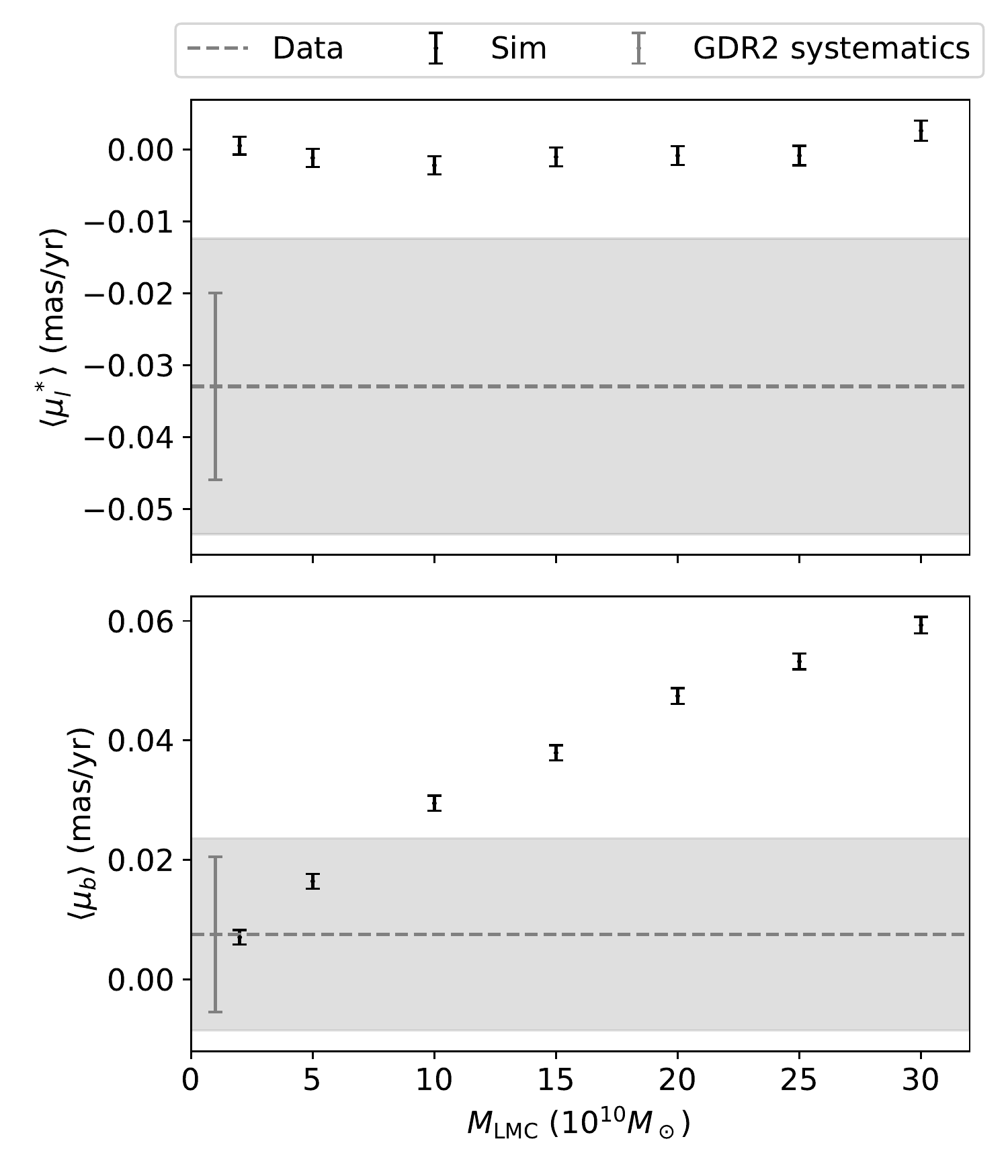}
\caption{Comparison of the mean proper motions in observations and simulations. The \textit{top panel} and \textit{bottom panel} show the mean proper motion in Galactic longitude ($\langle \mu_l^* \rangle$) and latitude ($\langle \mu_b \rangle$) respectively. In each panel, the dashed line and shaded region show the observed mean proper motion and $1\sigma$ uncertainty. The grey error bar shows the systematic uncertainty in proper motion on large angular scales \protect\citep{Lindegren2018,Mignard2018}. The black error bars show the mean proper motion in simulated stellar haloes with different LMC masses. The observed proper motions are small and consistent with a broad range of LMC masses at the $2\sigma$ level.}  \label{fig:pm_vs_mlmc}
\end{figure}

Although the main focus of this work is on radial velocities, we also investigate the proper motions of the distant halo stars by cross-matching our sample with \textit{Gaia} DR2. Of the 492 stars in our sample which pass all cuts, 468 have proper motions. We compute the reflex corrected proper motions, $\mu_l^*, \mu_b$, and Monte Carlo the observed uncertainties in distance and proper motion (including covariance) 10,000 times to get their associated uncertainties. Since the signature of the LMC's infall on the Milky Way's stellar halo is predicted to be fairly uniform on the sky \citep[e.g.][]{Erkal2020}, we fit a Gaussian to each proper motion for the entire sample. We find $\langle \mu_l^* \rangle = -0.03 \pm 0.02$ mas/yr and $\langle \mu_b \rangle = 0.01 \pm 0.02$ mas/yr. In \Figref{fig:pm_vs_mlmc} we show how this compares with the expected proper motion signature due to different mass LMCs. Both proper motions are consistent with a broad range of LMC masses within the $2\sigma$ uncertainty. With better proper motions soon available in \textit{Gaia} EDR3, it will be interesting to see whether the proper motions can also be used to constrain the LMC mass.


\subsection{Rotating halo} \label{sec:rotating}

Next we consider whether the observed radial velocities are consistent with a rotating stellar halo. As in \cite{deason_etal_2017}, we model the halo as having a constant azimuthal velocity and an isotropic dispersion in velocity. We allow for an arbitrary rotation of the rotational axis. Thus our model has four parameters: two angles describing the rotation axis, the mean azimuthal velocity, and a velocity dispersion. For each set of these four parameters, we compute the kinematic observables (i.e. $v_{\rm GSR}, \mu_l^*, \mu_b$) at the location of each star in our sample. We note that when we compute these observables, we ignore the distance uncertainty for each star and thus have a unique set of predicted kinematic observables for each star and rotation model. First, we fit only the radial velocities to determine whether they are consistent with a rotating halo. For the likelihood we use a Gaussian based on the observed radial velocity and its associated uncertainty compared to our predicted radial velocity. We assume a uniform prior on the azimuthal angle between $(0,2\pi)$, a prior of $p(\phi) \propto \cos(\phi)$ for the polar angle for $\phi$ between $(-\pi/2,\pi/2)$, a uniform prior on the rotational velocity between $(0,500)$ km/s, and uniform prior on the velocity dispersion between $(0,500)$ km/s. We explore the likelihood surface with an MCMC using \textsc{emcee} \citep{emcee}. We use 10,000 steps, 10 walkers, and a burn-in of 5,000 steps. We find a good fit with a rotational velocity of $172^{+49}_{-43}$ km/s with a median rotation axis roughly pointed in the $+y$ direction. The mean radial velocities and proper motions from the best-fit model are shown in \Figref{fig:rotating_model}. We note that this rotational velocity is comparable to the circular velocity of the Milky Way halo beyond 50 kpc \citep[e.g.][]{mcmillan_MW_2017,sgr_tango} and is thus quite unlikely.

Furthermore, such a rapidly rotating stellar halo would also possess significant proper motions. For comparison with the data, we compute the mean proper motions, $\langle \mu_l^* \rangle, \langle \mu_b \rangle$, at the location of the stars in our sample which have proper motions in \textit{Gaia} DR2 for 1,000 draws of our posterior chains. We find $(\langle \mu_l^* \rangle, \langle \mu_b \rangle) = (0.13, 0.20)$ mas/yr. These are ruled out by the observed mean proper motions discussed in \Secref{sec:pms} which are significantly smaller (see \figref{fig:pm_vs_mlmc}). 

Finally, we also attempt to simultaneously fit the observed radial velocities and proper motions with a rotating halo model which gives a halo with a rotational velocity of $18.5^{+5.1}_{-5.7}$ km/s. This slow of a halo rotation results in negligible mean radial velocities in the North and South, $+1.6$ km/s and $-1.7$ km/s respectively, as well as modest mean proper motions of $(\langle \mu_l^* \rangle, \langle \mu_b \rangle) = (-0.01,0.02)$ mas/yr. Thus, while a rapidly rotating halo is fairly consistent with the observed radial velocities, this implies that the observed velocity signature cannot be explained by pure axisymmetric rotation.

\subsection{Stellar halo substructure} \label{sec:substructure}

Although the radial velocity and proper motion of the distant Milky Way stellar halo are consistent with the predicted effect of the LMC, another possibility is that the observed velocity shifts are due to substructure in the stellar halo. Such a substructure would need to span a large portion of the sky encompassed by our sample and have a similar velocity dispersion to the stellar halo since we do not see any significant features with a low velocity dispersion in \Figref{fig:vgsr_dist}. One possible candidate for such a broad debris field could be tidal shells from the Gaia-Enceladus-Sausage \citep[GES,][]{sausage_disc,ge_disc} merger with our Galaxy. These shells should exhibit clear caustics in the space of radius versus radial velocity, as well as pile-ups near the apocenter of the shell \citep[e.g.][]{Sanderson_Helmi_2013}. Although we do not see any strong evidence for such features in this sample (see Fig. 3 and associated discussion in Deason et al. in prep.), larger upcoming spectroscopic surveys such as WEAVE, DESI, and 4MOST will be able to better explore this possibility. 

Furthermore, we note that there does appear to be some small-scale substructure in our sample. In the Southern hemisphere, the slight excess of stars with $0<v_{\rm GSR} < 50$ km/s see in \Figref{fig:vgsr_dist} are responsible for the patch of less negative radial velocities seen in \Figref{fig:vgsr_on_sky}. We note that if these stars are due to substructure, their removal would make the mean velocity in the South even more negative.

\subsection{Uncertainty in the Milky Way mass}

Throughout this work we have kept the Milky Way mass fixed and only varied the mass of the LMC. However, given that there is substantial uncertainty in the Milky Way mass \citep[e.g.][]{Wang_etal_2020}, we investigate what happens if we increase our Milky Way mass. For this test, we keep the Milky Way disk and bulge unchanged, but increase the Milky Way halo mass by 50\% to $1.2\times10^{12} M_\odot$. Since we keep the same scale radius and concentration for the halo, we note that this means the inner Milky Way will be inconsistent with measurements. We rerun the simulation described in \Secref{sec:sims} with this more massive Milky Way in the presence of a $1.5\times10^{11} M_\odot$ LMC, make mock observations of the resulting halo, and compute the mean radial velocity in the North and South. Interestingly, we find that increasing the halo mass by 50\% only changes the predicted mean radial velocity by $\sim15\%$ in the Southern hemisphere. This modest effect is likely due to the fact that as the LMC falls in, the part of the Milky Way which is decoupling from the outer Galaxy (i.e. within $\sim 30$ kpc), has a substantial mass contribution from the Milky Way disk. If we kept the same mass profile within this region, it is likely that the change in predicted radial velocity would be even smaller. Thus, the predicted shift in velocity is only weakly dependent on the total Milky Way mass. 

\subsection{Stellar halo sloshing as a tracer of past interactions}

Interestingly, the disk and stellar halo in Andromeda show a significant velocity offset \citep[see Fig. 7 in ][]{gilbert_etal_2018}. This may be due to the accretion of a massive satellite and indeed \cite{hammer_etal_2018} suggest Andromeda may have experienced a 4:1 merger which finished 2-3 Gyr ago. In light of this we note that the stellar halo sloshing seen in the Milky Way is likely long lived. The orbital periods of stars at 30 kpc (i.e. the region within which the halo responded adiabatically) are $\sim 1$ Gyr \citep[e.g. Fig. 11 in ][]{orphan_modelling} and it will likely take several orbital periods for the Milky Way stellar halo to reequilibrate after the LMC's perturbation. Thus, a velocity offset of the disk and stellar halo in external galaxies may be a useful tracer of substantial mergers. We note that it is possible that there could still be some residual sloshing in the Milky Way due to the GES merger \citep{sausage_disc,ge_disc}. However, given how well the predicted LMC signal matches the observed velocity structure of the outer halo, and given the large LMC masses inferred from other techniques \citep[e.g.][]{kallivayalil_etal_2013,penarrubia_lmc,orphan_modelling,sgr_tango}, we do not believe that the GES is responsible.  More work is needed with $N$-body simulations to investigate how long this sloshing persists.

\section{Conclusions} \label{sec:conclusions}

As predicted by a number of works \citep[e.g.][]{orphan_modelling,lmc_wake,petersen_penarrubia,Erkal2020,Garavito-Camargo_etal_2020}, we show for the first time that we have a substantial motion with respect to the outer parts of the Milky Way stellar halo\footnote{We note that in the final phases of preparing this manuscript we became aware of the work of Petersen \& Pe{\~n}arrubia in prep. which has independently discovered the same bulk motion in the outer stellar halo.}. This motion is visible in the mean radial velocity of stars which are redshifted (blueshifted) in the Northern (Southern) hemisphere, consistent with our moving mostly downwards with respect to the outer halo. The observed radial velocity signature is well described by the effect of an LMC with a mass of $1.5\times10^{11} M_\odot$. The observed proper motion signal is small and consistent with a broad range of LMC masses at the $2\sigma$ level. We will explore this further with improved proper motions in \textit{Gaia} EDR3.

This confirmation that the inner Milky Way is ``sloshing" about with respect to the outer stellar halo has a number of important implications:

\begin{itemize}
    \item Tracers in the outskirts of the Milky  (beyond 50 kpc) are not in equilibrium. As explained in \cite{Erkal2020}, if this disequilibrium is not accounted for, estimates of the Milky Way mass will be biased in the outer halo. We perform this correction and measure the Milky Way mass out to 100 kpc in a companion work (Deason et al. in prep.).
    \item This is another piece of evidence which confirms that the LMC has had a substantial effect on the Milky Way, consistent with it being on a first approach to our Galaxy and having retained a significant amount of dark matter.
    \item It confirms that the inner Milky Way is not an inertial reference frame but instead that we have been substantially accelerated by the LMC over the past 2 Gyr \citep[e.g. see Fig. 11 of][]{orphan_modelling}. This motion must be accounted for to accurately model the orbits of satellites and dwarf galaxies in the distant Milky Way halo \citep[e.g.][]{lmc_sats}. Furthermore, this motion has a larger effect on the trajectory of hypervelocity stars than the deflection expected from triaxial haloes in $\Lambda$CDM \citep{Boubert_etal_2020}.
    \item  Given the results of recent numerical simulations, this result implies that the dark matter halo of both the Milky Way and the LMC have probably also been significantly deformed \citep[e.g.][]{lmc_wake,petersen_penarrubia,Garavito-Camargo_etal_2020}. If this can be tested and verified, either with the stellar halo or stellar streams in the Southern hemisphere \citep[e.g.][]{li_s5}, it would be a stunning confirmation of the dark matter paradigm.
\end{itemize}

Future spectroscopic surveys like WEAVE, 4MOST, and DESI, combined with updated proper motions from \textit{Gaia}, will allow for a much better characterization of the LMC's effect on the Milky Way, allowing us to better understand both our Galaxy and the LMC.

\section*{Data availability}

The present day snapshot of the fiducial simulation with a $1.5\times10^{11} M_\odot$ LMC is publicly available at \url{https://doi.org/10.5281/zenodo.3630283}. The other snapshots will be made available upon request. All of the data used in this work, apart from the K-giants from \cite{yang_etal_2019}, are publicly available. These stars will be made available in Xue et al. in prep. Our final catalog for the publicly available stars, including cross-match to \textit{Gaia} DR2 and Sagittarius membership, is available upon request. 

\section*{Acknowledgements}

We thank the key workers who made this research possible, especially during the COVID-19 pandemic. We thank Eugene Vasiliev for fruitful discussions and for help with using \textsc{agama}. AD is supported by a Royal Society University Research Fellowship, the Leverhulme Trust, and by the Science and Technology Facilities Council (STFC) [grant numbers ST/P000541/1, ST/T000244/1]. X.-X. Xue, C. Liu, G. Zhao and L. Zhang acknowledge support from the National Natural Science Foundation of China under grants Nos. 11988101, 11873052, 11873057,11890694, 11773033 and National Key R\&D Program of China No. 2019YFA0405500. 

This research made use of \textsc{ipython} \citep{IPython}, python packages \textsc{numpy} \citep{numpy}, \textsc{matplotlib} \citep{matplotlib}, and \textsc{scipy} \citep{scipy}. This research also made use of Astropy,\footnote{http://www.astropy.org} a community-developed core Python package for Astronomy \citep{astropy:2013, astropy:2018}.

\bibliographystyle{mn2e_long}
\bibliography{citations_lmc}

\appendix

\section{Rotating halo}

\begin{figure*}
\centering
\includegraphics[width=0.99\textwidth]{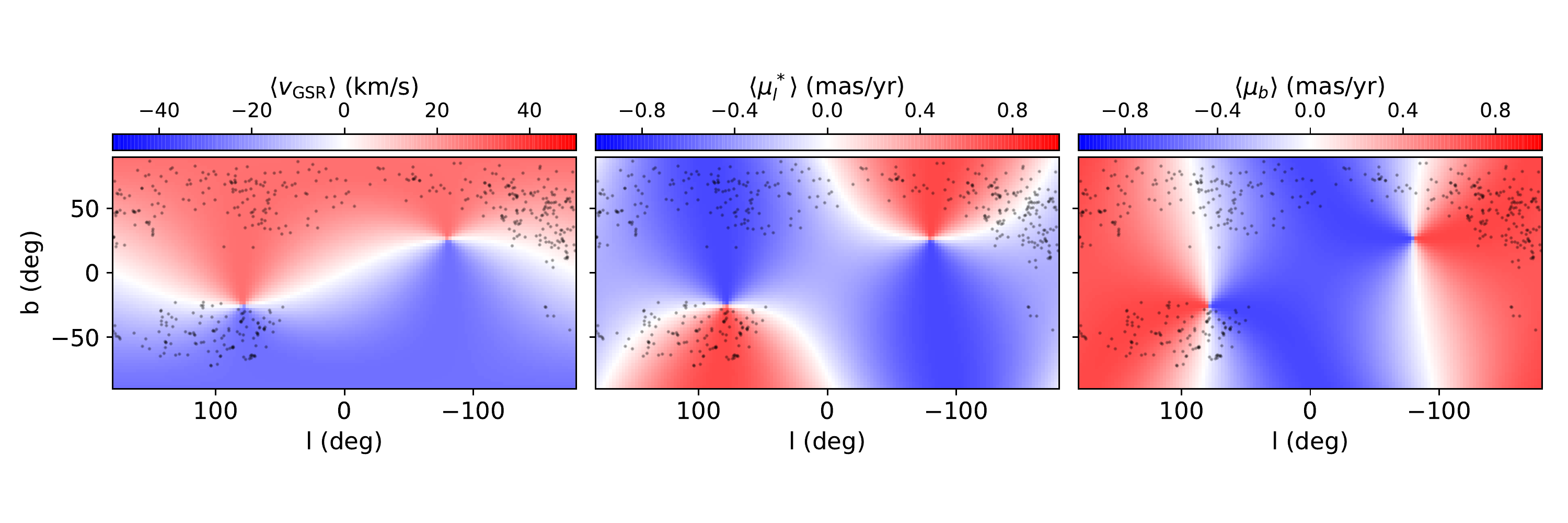}
\caption{Best-fit rotating halo model fit to the observed radial velocities. The left, middle, and right panels show the mean radial velocity, mean proper motion in $l$, and mean proper motion in $b$ respectively. The means are computed in $2.5^\circ$x$2.5^\circ$ pixels on the sky assuming a Galactocentric distance of $50$ kpc. In each panel, the black points correspond to the location of the stars in our sample. While the radial velocity pattern on the sky is similar to observations (see \protect\figref{fig:vgsr_on_sky}), this rotating model is ruled out by the substantial proper motions. Note that to aid in comparison, we have used the same colorbar for the left panel as in the left panel of \protect\figref{fig:vgsr_on_sky}.}  \label{fig:rotating_model}
\end{figure*}

In \Figref{fig:rotating_model} we show the best-fit rotating model fit to the radial velocities described in \Secref{sec:rotating}. This model has a rotational velocity of $\sim186$ km/s and a rotational axis $\sim32^\circ$ below the plane of the Milky Way disk. We compute the mean radial velocity and reflex corrected proper motions in $l,b$ in $2.5^\circ$x$2.5^\circ$ pixels on the sky assuming a Galactocentric distance of 50 kpc. Overall, this model provides a good fit to the pattern of radial velocities on the sky (compare with left panel of \figref{fig:vgsr_on_sky}) although we note that the observed radial velocities are substantially more negative in the South around $(l,b) \sim (80^\circ,-50^\circ)$. Despite this good agreement with the radial velocities, this model predicts substantial proper motions that are ruled out by observations in \textit{Gaia} DR2.

\end{document}